\documentclass[draft]{agujournal}

\draftfalse

\usepackage[utf8]{inputenc} 
\usepackage[T1]{fontenc}      
\usepackage[english]{babel}
\usepackage{graphicx}
\usepackage[a4paper]{geometry}
\usepackage{hyperref}
\usepackage{makeidx}
\usepackage{amsmath,amsfonts,amssymb}
\usepackage{footnote}
\usepackage{fancyhdr}
\usepackage{moreverb}
\usepackage{here}
\usepackage{gensymb}
\usepackage{amssymb}
\usepackage{natbib}
\usepackage{ulem}
\usepackage{caption}
\bibpunct{(}{)}{;}{a}{}{,}

\usepackage[colorinlistoftodos]{todonotes}

\newcommand{\reff}{Fig.\ref}
\newcommand{\refs}{Sec.\ref}
\newcommand{\refe}{Eq.\ref}
\newcommand{\reft}{Tab.\ref}

\usepackage{color}

\journalname{Geophysical Research Letters}

\begin{document}

\title{Simulation of plasmaspheric plume impact on dayside magnetic reconnection}




\authors{J. Dargent\affil{1}, 
N. Aunai \affil{2}, 
B. Lavraud \affil{3}, S. Toledo-Redondo \affil{3,4}, F. Califano \affil{1}}

\affiliation{1}{Dipartimento di Fisica "E. Fermi", Universit\'a di Pisa, Pisa, Italy}
\affiliation{2}{LPP, CNRS, Ecole polytechnique, UPMC Univ Paris 06, Univ.  Paris-Sud, Observatoire de Paris, Universit\'e Paris-Saclay, Sorbonne Universit\'es, PSL Research University, Palaiseau, France}
\affiliation{3}{Institut de Recherche en Astrophysique et Plan\'etologie, Universit\'e de Toulouse, CNRS, UPS, CNES,
Toulouse, France}
\affiliation{4}{Department of Electromagnetism and Electronics, University of Murcia, Murcia, Spain}

\author{Nicolas Aunai}

\author{Beno\^it Lavraud}

\author{Sergio Toledo-Redondo}

\correspondingauthor{J\'er\'emy Dargent}{jeremy.dargent@df.unipi.it}

\date{\today}

\begin{keypoints}
\item We perform Particle-In-Cell simulation of asymmetric magnetic reconnection including the impact of cold plasmaspheric plume
\item Inclusion of the plume reduces the reconnection rate, the reduction is explained by the additional mass in the system (mass-loading effect)
\item The temperature of the plume does not influence the reconnection rate
\end{keypoints}

\begin{abstract}

During periods of strong magnetic activity, cold dense plasma from the plasmasphere typically forms a plume extending towards the dayside magnetopause, eventually reaching it. 
In this work, we present a large-scale two-dimensional fully kinetic Particle-In-Cell simulation of a reconnecting magnetopause hit by a propagating plasmaspheric plume. 
The simulation is designed so that it undergoes four distinct phases: initial unsteady state, steady state prior to plume arrival at the magnetospause, plume interaction and steady state once the plume is well engulfed in the reconnection site.
We show the evolution of the magnetopause's dynamics subjected to the modification of the inflowing plasma.
Our main result is that the change in the plasma temperature (cold protons in the plume) have no effects on the magnetic reconnection rate, which on average depends only on the inflowing magnetic field and total ion density, before, during and after the impact.

\end{abstract}

\section{Introduction}
\label{sec:intro}

During strong magnetic activity periods cold dense plasma emerging from the plasmasphere typically generates a plume extending towards, and eventually reaching, the dayside magnetopause.
The density of these plumes is comparable or even larger than the magnetosheath densities \citep{McFadden2008,Walsh2014}, and therefore mass-loads the reconnection site and reduces the characteristic Alfvén speed.
As a result, the local reconnection rate, i.e. the amount of magnetic flux reconnected per unit of time is reduced and magnetic reconnection becomes less efficient at converting magnetic energy into kinetic energy of the particles \citep{Borovsky2006,Borovsky2007,Walsh2013}.
So far, numerical studies have relied on Magnetohydrodynamics (MHD) to model the effect of plasmaspheric plumes on magnetopause reconnection \citep{Borovsky2007,Borovsky2008,Ouellette2016}.
However, accounting for kinetic scales processes is known to be important for reconnection modeling.
Fully kinetic simulations of magnetic reconnection (even without a plume) are small and even the largest so far barely reach a steady state at scales at which ions are fully frozen in the magnetic field \citep{Malakit2013,Dargent2017}.
Consequently, plume simulations until now have been unable to model the kinetic dynamics of magnetic reconnection while kinetic simulations of magnetic reconnection have yet to reach scales relevant with fluid dynamics.
We can therefore wonder to what extent kinetic solutions differ from the fluid ones, in particular in the context of the impact of a cold plasma plume.

\citet{Cassak2007} proposed a MHD scaling law of the asymmetric magnetic reconnection rate $R$ in steady state depending only on the inflowing plasma density $n$ and magnetic field $B$ values: 
\begin{eqnarray}
	R &\sim& \frac{B_1 B_2}{B_1+B_2} v_{out} \frac{2\delta}{L}
    \label{eq:CSR} \\
	v_{out} &=& \sqrt{B_1 B_2 \frac{B_1+B_2}{B_1 n_2 + B_2 n_1} }
\end{eqnarray}
where the subscripts 1 and 2 refer to the two sides of the layer and $\delta/L$ is the aspect ratio of the diffusion region.
This model proved to be quite accurate \citep{Birn2008,Borovsky2008}. 
However, the domain of validity of this model is limited. 
It only gives the local reconnection rate calculated with parameters near the X point.
On larger scales, the global reconnection rate can be calculated by the net force acting on the flow in the magnetosheath.
Such a global reconnection rate often differ from the local reconnection rate \citep{Zhang2016}.
As a MHD model, \citet{Cassak2007}  neglect all kinetic processes potentially impacting magnetic reconnection \citep{Hesse2013,Dargent2017,Tenfjord2019,Kolsto2020}.
Finally, due to the steady state assumption, one can wonder if this model is still applicable in during variations of the external environment, such as during the impact of a plasmaspheric plume.

Observation studies \citep{Toledo2015,Andre2016} suggest that the presence of magnetized cold ions could reduce the current density, thus the Hall electric field, and consequently could have an impact on the reconnection rate, although in-situ spacecraft observations cannot measure the reconnection rate accurately \citep{Genestreti2018}.
However, \citet{Toledo2018} showed, using Particle-In-Cell simulations, that even if the electric field is locally reduced by cold ions, the potential drop averaged through the current layer, and therefore the mean reconnection electric field, remains unaffected.
More generally, recent kinetic simulations of magnetic reconnection with cold ions suggest that the effect of cold ions on the reconnection rate is negligible for both symmetric \citep{Divin2016} and asymmetric \citep{Dargent2017} layers. 
Those works, however, consider cases where cold ions are a minority species (33\% of the magnetosphere, itself three times less dense than the magnetosheath in \citet{Dargent2017,Dargent2019}) unable to modify significantly the reconnection layer dynamics such as, instead, in the case of a plume impact.

In the work presented here we have performed a fully kinetic simulation to investigate the impact of a plasmaspheric plume with a reconnecting dayside magnetopause.
We made the domain large enough to include the transition from kinetic scale to frozen-in scales and long enough to capture the modifications caused by a plasmasperic plume on a fully developed exhaust.
The results presented in this paper especially focus on the reconnection rate evolution depending on the inflowing plasma parameters.

\section{Simulation setup}
\label{sec:setup}

In this paper, we present a two-dimensional (2D-3V) fully kinetic simulation of the impact of a large density plasmaspheric plume on ongoing asymmetric magnetic reconnection using the Particle-In-Cell (PIC) code SMILEI \citep{Derouillat2017}.
The plume is modeled in the simulation by a large amount of cold plasma (twice as large as the magnetosheath density itself, see \reft{tab:initasym}) initially located in the magnetosphere and at some distance from the current sheet.
Such dense plume is probably unfrequent but nevertheless it has already been observed \citep{Walsh2014} and allows us to magnify the effects driven by the plume impact.
This plume is then advected  towards the reconnection site by the reconnection inflow. 
We can define four main phases of the simulation.
The first one (phase I) corresponds to the initial growth of the reconnection rate occuring between the pristine magnetosphere and the magnetosheath, before reaching a steady state.
In the second one (phase II) we observe the quasi-steady magnetic reconnection without cold ions since the plume has not yet reached the current sheet.
The third phase (phase III) corresponds to the transition period when the plume impacts the reconnection layer.
The last phase (phase IV) is that of a relatively steady state reconnection with a very dense plume located at the inflow region.

All physical quantities here are normalized using ion characteristic quantities.
The magnetic field and density are normalized to the values in the magnetosheath, i.e. $B_0$ and $n_0$, respectively.
The masses and charges are normalized to the proton mass $m_p$ and charge $e$,
time is normalized to the inverse of the proton gyrofrequency $\omega_{ci}^{-1}=m_p/eB_0$ and
length to the proton inertial length $d_i=c/\omega_{pi}$, where $c$ is the speed of light in vacuum and $ \omega_{pi}=\sqrt{n_0e^2/m_p\epsilon_0}$ is the proton plasma frequency.
Velocities are normalized to the Alfv\'en velocity $v_{Al} = d_i \omega_{ci}$.

The initial condition consists in one current layer varying in the $y$-direction and lying in the $(x,y)$ plane.
The initial fluid equilibrium is perturbed in order to trigger magnetic reconnection (see in Appendix).
The domain has size given by  $(x_{max},y_{max})=(1280,256)d_i$.
There are $n_x=25600$ cells in the $x$ direction, $n_y=10240$ cells in the $y$ direction and initially $50$ particles per cell per population. 
Plasma moments and electromagnetic forces are calculated using second order interpolation.
Particles are loaded using local Maxwellian distributions.
The time step is $dt=8.4 \cdot 10^{-4} \omega_{ci}^{-1}$. 
The total simulation time is $T=800 \omega_{ci}^{-1}$. 
The mass ratio $m_i/m_e$ is 25.
We fix $\omega_{pe}/\omega_{ce}=4$, i.e. $c/v_{Al} = 20$.
The system has periodic boundary conditions in the $x$ direction and reflective boundary conditions in the $y$ direction.
The current layer is located at $y_0=y_{max}/2=128~d_i$.
The plume is initially located at $\Delta y=20~d_i$ away from the current layer on the magnetospheric side, i.e. at a position $y_p=108~d_i$.
The value of $\Delta y$ is determined by using the \citet{Cassak2007} formula giving a characteristic time $t_{imp}$ for the plume to reach the current layer is approximately 300 $\omega_{ci}^{-1}$.

The asymptotic magnetic field value and the temperatures and density for each population in the different area are summarized in \reft{tab:initasym}.
From these values, we can estimate that the Alfv\'en velocity of hot ions would of $\sim6.3~v_{Al}$. 
They are therefore able to travel the whole box in $\sim200~\omega_{ci}^{-1}$.
However, given the low density of heavy ions and the fact that magnetic reconnection is already well developed at that time, we consider that this should not impact the dynamics.
The calculations of the profiles of density, temperature and magnetic fields are presented in the Appendix.

\begin{table}
	\begin{center}
   \begin{tabular}{ | l || c | c | c | c | c | c | c | c | c | c | c | c | }
     \hline
     Quantities    & $B$ & $n$ & $T_i$ & $n_{ish}$ & $n_{ih}$ & $n_{ip}$ & $T_{ish}$ & $T_{ih}$ & $T_{ip}$ & $\beta_{ish}$ & $\beta_{ih}$ & $\beta_{ip}$ \\ \hline
     Magnetosheath & -1 &  1  & 2.9 & 1 & 0 & 0 & 2.9 & 0 & 0 & 5.8 & 0 & 0 \\ \hline
     Magnetosphere (no plume) &  2 & 0.1 & 16.7 & 0 & 0.1 & 0 & 0 & 16.7 & 0 & 0 & 0.84 & 0 \\ \hline
     Magnetosphere (with plume) &  2 & 2.096 & 0.8 & 0 & 0.096 & 2 & 0 & 16.7 & 0.03 & 0 & 0.80 & 0.03 \\ 
     \hline
   \end{tabular}
 	\end{center}
   \caption{Asymptotic values of the different quantities normalized by ion scale quantities and the resulting plasma $\beta$.}
   \label{tab:initasym}
\end{table}

\section{Simulation results}

\subsection{Overview}

We aim at studying the impact of a plasmaspheric plume on magnetic reconnection at the dayside magnetopause.
This study includes the propagation of the (cold) plume ions in the exhaust and their impact on the structure of the exhaust.
The inflowing cold ions do not only affect the exhaust through the X line but also through the magnetosperic separatrices because of their drift there \citep{Dargent2019}.
Thus, to fully capture the impact on a plasmaspheric plume on magnetic reconnection, we need well-developed exhausts in both phases with and without the plume.
For that purpose we run the simulation until reaching a quasi steady state with exhausts of plasmaspheric plume plasma of the order of $ 100~d_i$ on both sides.
Phase I lasts from $t=0~\omega_{ci}^{-1}$ to $t\approx50~\omega_{ci}^{-1}$.
In phase II, the plume is still far from the magnetopause and the reconnected plasmas are essentially the tenuous hot magnetospheric plasma and the denser magnetosheath plasma.
This phase is characterized by the transient formation of plasmoids, which have a large impact on the local reconnection rate (see \refs{sec:RR}).
This phase lasts until the plume arrives at $t\approx300~\omega_{ci}^{-1}$.
It lasts long enough for the exhausts to develop for more than $100~d_i$ from either sides of the diffusion region.
The impact of the plume determines the beginning of phase III, which is a transition period.
This period will be further described in \refs{sec:RR}.
After this transition, phase IV is characterized by quasi-steady magnetic reconnection in the presence of the plume.
This period starts before $t=400~\omega_{ci}^{-1}$ and lasts until the end of the simulation at $t=800~\omega_{ci}^{-1}$.
This phase is characterized by the formation of two big plasmoids.
This long time allows for the development of a long exhaust on either side of the diffusion region, despite the decrease of the Alfv\'en velocity due to the mass loading effect by the plume.

\begin{figure}[h!]
  \includegraphics[width=12.5cm]{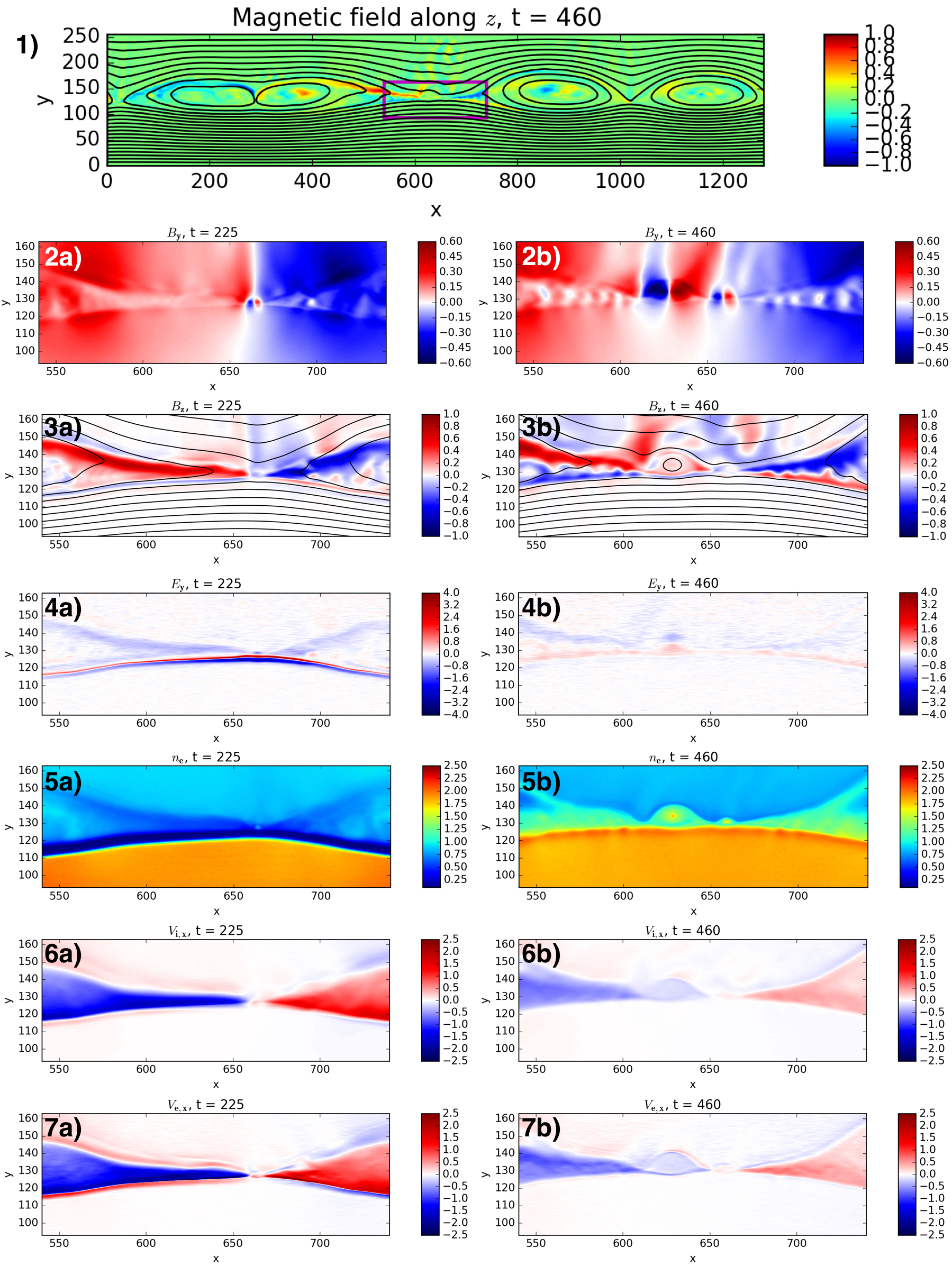}
  \caption{
 1) - Out-of-plane magnetic field $B_z$ in the whole box.
  In-plane magnetic field lines are depicted as thick black lines.
  The purple square shows the zoomed area used for all the other pictures of this figure.
 Simulation fields before ($a$) and after ($b$) the impact of the plume:
 2) - Magnetic field $B_y$.
 3) - Magnetic field $B_z$. In-plane magnetic field lines are depicted as thick black lines.
 4) - Electric field $E_y$.
 5) - Electron density.
 6) - Ion velocity $V_{i,x}$.
 7) - Electron velocity $V_{e,x}$.
  } 
  \label{fig:fields}
\end{figure}

In \reff{fig:fields} we show the shaded iso-contours of various quantities at time $t=225$ (left panels, phase II) and $t=460$ (first frame and right panels, phase IV), i.e. before and after the impact of the plume.
In the top frame we show the magnetic field in the whole box at $t=460~\omega_{ci}^{-1}$ (out-of-plane component as colors and in-plane components as black lines).
The magnetic large-scale configuration we see here results from an inverse cascade process of the initial many island structures emerging due to the tearing instability during phase I. 
We observe the formation of about seven-eight magnetic small islands (not shown here), corresponding to the growth of the most unstable modes. These islands then start to interact and merge very efficiently because of the 2D geometry \citep{Malara1992}.
The purple box in \reff{fig:fields}.$1$ marks the region plotted in the following panels, $2$ to $7$.
\reff{fig:fields}.$2$ we show the $y$-component of the magnetic field which displays a clear bipolar signature of plasmoids. 
Plasmoids can however be observed on other quantities.
Furthermore, a local zoom at $t=460$ on the reconnection layer (not shown here) shows that the in-plane magnetic field starts to bend and flap inside the exhaust region. 
Such curving of the field lines is also visible on the shaded iso-contours of $B_z$ in \reff{fig:fields}.$3b$.
We conjecture that such fluctuations result from the development of a  fire-hose like instability driven by local anisotropy. 
This point is not our objective here and will be matter for future work.
\reff{fig:fields}.$3$ shows the out-of-plane magnetic field $B_z$ and its quadrupolar Hall structure together with the in-plane field lines overplotted in black.
We observe that the Hall structure changes together with the density asymmetry evolution between $t=215$ and $t=460$, \reff{fig:fields}.$3a$ and $3b$. 
Indeed, the strong density asymmetry before the impact of the plume makes the Hall fields peaking on the magnetosheath side of the exhaust (\reff{fig:fields}.$3a$), whereas it becomes more quadrupolar as the plume fills the exhaust (\reff{fig:fields}.$3b$).
We note that this occurs even though the magnetic asymmetry remains unchanged during this interval.
\reff{fig:fields}.$4$ shows the electric field $E_y$ and its characteristic Hall bipolar structure.
The asymmetry density evolution driven by the plume entry also produces a strong decrease of the Hall electric field amplitude (see \reff{fig:fields}.$4b$) because of the density increase \citep{Toledo2018}, pushing the system towards a more symmetric configuration.
The plasma density is shown in \reff{fig:fields}.$5$.
In particular in \reff{fig:fields}.$5a$ we distinguish the tenuous magnetospheric plasma before the impact of the plume (dark blue) reconnecting with the magnetosheath plasma (light blue).
The dense plasma of the plume (orange) is arriving from below. 
In \reff{fig:fields}.$5b$, it is this plasma which is reconnecting. Finally, \reff{fig:fields}.$6$ (\reff{fig:fields}.$7$) shows the ions (electrons) exhaust velocity $V_x$.
These two quantities look very similar except at the X-point and along the exhaust boundaries.
We also observe that the exhaust velocity slows down as soon as the plume impacts on the reconnection region, as expected because of a mass loading effect \citep{Cassak2007,Borovsky2008,Walsh2013}.

\subsection{Reconnection rate dependencies}
\label{sec:RR}

We will now focus on the evolution of the reconnection rate during the whole simulation and compare it with the theoretical model by \citep{Cassak2007}.
From \refe{eq:CSR} we get a reconnection rate $R$ scaling with the inflowing density and magnetic field as follows:
\begin{equation}
	R \propto \frac{B_1 B_2}{B_1+B_2} v_{out} = B_1 B_2 \sqrt{\frac{B_1 B_2}{B_1+B_2} \frac{1}{B_1 n_2 + B_2 n_1} } = R_{CS}
	\label{eq:CS}
\end{equation}
where $B$ and $n$ are the asymptotic values of the norm of the magnetic field and density, respectively. 
The subscripts $1$ and $2$ indicate the two sides of the layer (in our case the magnetosphere and the magnetosheath, respectively).
At steady state, the quantity $R_{CS}$ can be seen as a normalization factor of the reconnection rate.
From \refe{eq:CSR} and \ref{eq:CS}, we get that the normalized reconnection rate is proportional to the aspect ratio:
\begin{equation}
	R' = \frac{R}{v_{out}B_{mean}} = \frac{R}{R_{CS}} \sim \frac{2 \delta}{L}
	\label{eq:Rp}
\end{equation}
\begin{figure}[h!]
  \includegraphics[width=14cm]{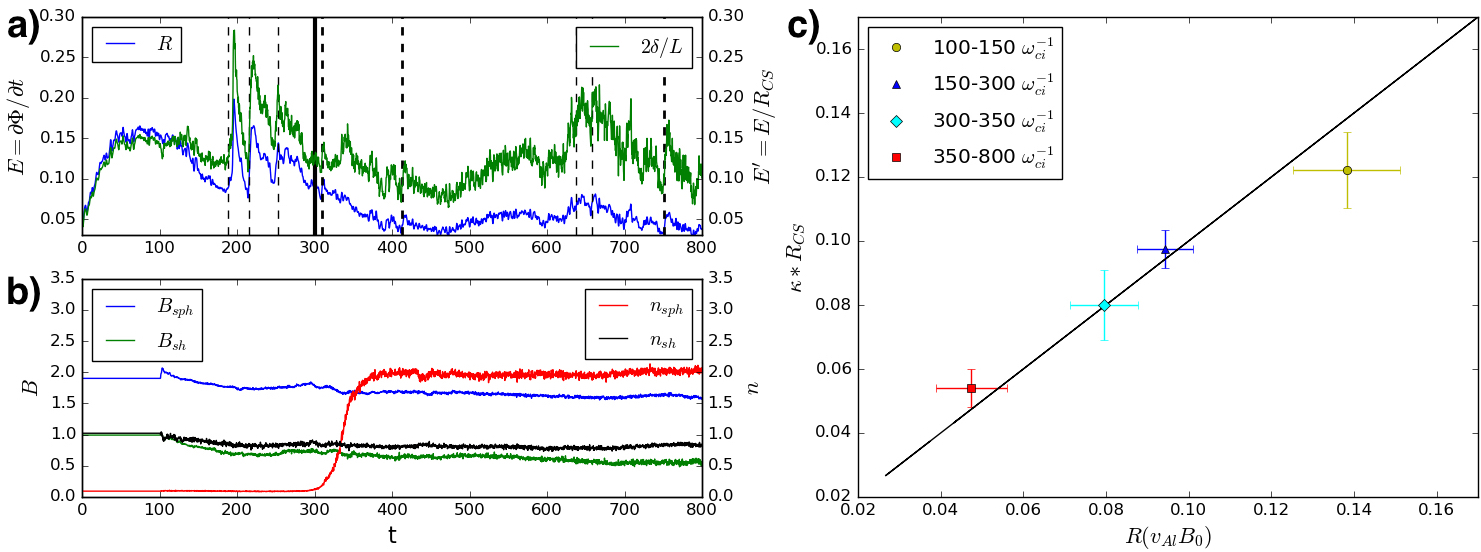}
  \caption{
  a - Reconnection rate $R$ of the plume simulation, in blue, and this same rate normalized by $R_{CS}$ (see \refe{eq:CS}), in green.
  Reconnection rate is here defined as the time derivative at the X point of the magnetic flux $\Phi$ in the simulation plane $(x,y)$.  This is equivalent to the out-of-plane electric field $E_z$ at the X point position, see \citet{Shay2001} or \citet{Pritchett2008} for a developed description of the magnetic flux and how reconnection rate comes from it.
  The vertical straight line indicate the impact time of the plasmaspheric plume.
  The vertical dashed lines show the formation times of plasmoids. 
  The thick ones are for the large plasmoids, which survive for hundreds of $\omega_{ci}^{-1}$, while the thin ones are for transient plasmoids.
  b - Asymptotic values of magnetic field ($B$) and density ($n$) used to normalized the reconnection rate.
  These values are taken at a distance of $\delta y =10~d_i$ (resp. $\delta y =20~d_i$) afar from the X point in the magnetosphere (resp. the magnetosheath) and is then shifted in time depending of the speed of convection of the plasma.
  c - Scatter plot, for all the times in the simulations such as $t>100~\omega_{ci}^{-1}$, of the reconnection rate $R$ versus $\kappa R_{CS}$, where $\kappa=0.127$ is a constant such as $\kappa R_{CS}/R$ scales along a slope of 1 (black curve, see the text).
  Each point correspond to the mean value of the reconnection rate on a time interval and the bars associated with them provide one standard deviation.
  }
  \label{fig:rates}
\end{figure}

In \reff{fig:rates}$a$, we plot the reconnection rate $R$ and the normalized reconnection rate $R' \sim 2 \delta/L$, blue and green line, respectively. 
In particular, the blue curve helps us in following the global dynamics of the simulation. 
In the initial phase up to $t\approx 50~\omega_{ci}^{-1}$, magnetic reconnection develops.
Then, the reconnection rate reaches a maximum and decreases between $t\approx 50~\omega_{ci}^{-1}$ and $t\approx 150~\omega_{ci}^{-1}$ during the overshoot phase (early phase II).
The overshoot is a typical feature observed in numerical simulations of magnetic reconnection and depends on the initial current sheet thickness \citep{Shay2007} and on the initial perturbation.
The time interval between $t\approx150~\omega_{ci}^{-1}$ and $t\approx300~\omega_{ci}^{-1}$ is the phase of quasi-steady magnetic reconnection without the plume (late phase II).
This phase is strongly affected by the formation of transient plasmoids, that impact the aspect ratio $\delta/L$ leading to large variations of the reconnection rate.
At $t\approx300~\omega_{ci}^{-1}$, the plasmaspheric plume impacts on the magnetopause and the reconnection rate decreases (phase III).
After $t\approx350~\omega_{ci}^{-1}$, the reconnection is once again in a quasi-steady state (phase IV).

The green curve in \reff{fig:rates}$a$ is obtained by dividing the reconnection rate $R$ by $R_{CS}$ (see \refe{eq:Rp}).
To calculate $R_{CS}$, we use the inflow plasma asymptotic values of density and magnetic field plotted in \reff{fig:rates}$b$. 
These values are taken in the magnetosphere (resp. the magnetosheath) at a distance of $\delta y =10~d_i$ (resp. $\delta y =20~d_i$) away from the X point.
To take into account the plasma convection, we calculate $R_{CS}$ at $t$ with plasma values taken at $t-\delta t$, where $\delta t \sim B_x/E_z  \delta y$.
$E_z$ turns out to be of the order of 0.1 (as expected from \citet{Liu2017}) and we used the initial values listed from \reft{tab:initasym} for $B_x$.
However, the estimation of $\delta t$ is not a strict equality.
To find a usable empirical relation between $\delta t$ and $\delta y$, we looked how much time the plume takes to drift from its initial position to the magnetopause (traveled distance  of $\Delta y = 20~ d_i$ and impact at the X point at $t=300~\omega_{ci}^{-1}$).
We find $\delta t=5 B_x \delta y$.
For the $\delta y$ chosen in \reff{fig:rates}.$2b$, we obtain a time shift of $\delta t=100~\omega_{ci}^{-1}$ in both cases.
The main feature of this green curve is that, despite large deviations at small scales (mainly due to plasmoids), the model of \citet{Cassak2007} holds in magnitude even for our extreme conditions.
It is also worth noticing that the green curve deviates from the blue curve before the impact of the plume at $t=300~\omega_{ci}^{-1}$.
The reason is that the simulation box contains a finite amount of magnetic flux.
As the magnetic field is reconnected, the inflowing magnetic flux will be depleted and the field amplitude will decrease.
Such a decrease is usually neglected but given the size and the time length of this simulation, we observe a small decrease of the inflowing magnetic field amplitude, of the order of 20\% between the beginning and the end of the simulation (see \reff{fig:rates}$b$).
A secondary feature of the green curve is that its steady state value is in-between $0.1$ and $0.2$, which is consistent with the work of \citet{Liu2017,Liu2018} about the fluid scale constraints on the reconnection rate.

In \reff{fig:rates}$c$, we can see the reconnection rate $R$ in our simulation versus the \citet{Cassak2007} normalization factor $R_{CS}$ (see \refe{eq:CS}) for each phase of the simulation.
For each phase, we give the mean value (point) and one standard deviation (bars) of the reconnection rates.
The rates excluded from the calculation because of plasmoids are all the rates for times $t$ such as $190<t<300$ and $635<t<675$.
The $R_{CS}$ term is normalized by a constant $\kappa$ for the slope between  $R$ and $R_{CS}$ to be equal to 1.
To get $\kappa$, we made a linear regression on the reconnection rates.
We notice that $\kappa \sim 2\delta/L \sim 0.1$.
The global picture of \reff{fig:rates}$c$ is that, except in presence of transient plasmoids the \citet{Cassak2007} theory describes very well the variations of the reconnection rate.
In the details, during the overshoot period (yellow dot in \reff{fig:rates}), the reconnection rate is a bit higher than predicted by the theory, but this is expected \citep{Shay2007}, since the steady state is not yet reached.
The quasi-steady states with (red square) and without (blue triangle) the plume scales well with the slope of 1.
Regarding the transition (light blue diamond), we also observe that the rates scale very well with the theory.
Furthermore, whatever the phase of the simulation, $R/R_{CS} \sim 0.1$ \citep{Cassak2017}.

\section{Conclusions}

We showed that during the impact of a plasmaspheric plume modeled by adding a cold proton population to the ion distribution, the magnetic reconnection rate is only affected by its contribution to the density. 
The reconnection rate turns out to be in agreement with the \citet{Cassak2007} model before, during and after the plume impact.
This means that this model remains valid whatever the temperature of the populations composing the plasma.
On the other hand, the local reconnection rate is affected by plasmoids, which modify the aspect ratio of the diffusion region.

The \citet{Cassak2007} model reveals that the reconnection rate is much more sensitive to magnetic field changes rather than to density variations. By applying \refe{eq:CS} to our initial current layer, we get that an increase of magnetospheric density from 0.1 to 2 (i.e. impact of the plume) is equivalent to a decrease of the magnetospheric magnetic field from 2 $B_0$ to 1.2 $B_0$, in term of induced variations of the reconnection rate. 
Such magnetic field changes on the magnetospheric side of the magnetopause are quite common and depend mainly on the Solar wind dynamic pressure conditions. 
We expect a similar effect on magnetic reconnection in the presence of magnetosphere magnetic field depletion.

The simulation presented here addresses for the first time asymmetric magnetic reconnection with a large density plume down to electron kinetic scale.
We observe the formation of large-scale plasmoids driven by the development of small scale instabilities.
We also observe, even far from the X line, the presence of the Hall electric field along the separatrices embedded in a MHD-like exhaust ($\mathbf{v}_e \simeq \mathbf{v}_i$). 
Furthermore, we observe that the impact of the plume changes the structure of the whole system eventually leading to a more symmetric layer and to a reduction of the exhaust velocity.
Finally, this work highlights the strong impact of plasmoids on the local reconnection rate
since they modify the aspect ratio of the diffusion region.
In \refe{eq:CS} a constant aspect ratio is assumed, but the reconnection rate scales linearly with it.
A future study will focus on the formation of plasmoids in the diffusion region and their impact on magnetic reconnection.

\section*{Acknowledgments}

The simulation data needed for this study is included as supplementary informations.
This project (JD, FC) has received funding from the European Union's Horizon 2020 research and innovation programme under grant agreement No 776262 (AIDA).
The authors thank the SMILEI development team and especially Julien Dérouillat, Arnaud Beck, Frederic Perez, Tommaso Vinci and Michael Grech.
This work was performed using HPC resources from GENCI-TGCC (special grant t201604s020).
We acknowledge support from the ISSI international team Cold plasma of ionospheric origin in the Earth’s magnetosphere.
Research at IRAP was supported by CNRS, CNES and the University of Toulouse. 
STR acknowledges support of the of the Ministry of Economy and Competitiveness (MINECO) of Spain (grant FIS2017-90102-R).

\section*{Appendix}

The simulation is initialized with an electric field $\mathbf{E}$ null everywhere and a magnetic field $\mathbf{B}$:
\begin{equation}
	\textbf{B}(x,y) = \frac{1}{B_r} \left[-\tanh \left(\frac{y-y_0}{L} + arctanh \left(\frac{B_r-1}{B_r+1}\right)\right) \frac{B_r+1}{2} - \frac{B_r-1}{2} \right] \textbf{u}_x
	\label{eq:B}
\end{equation}
with $L=1$, $B_r=|B_{sheath}/B_{sphere}|$ the magnetic field ratio between both sides of the current sheet and $\textbf{u}_x$ the unit vector in the $x$ direction. 
We choose $B_r = 0.5$.

The total temperature $T=T_i+T_e$ is determined in order to preserve the pressure balance.
The electron to ion temperature ratio is constant and chosen equal to $\theta=T_e/T_i=0.2$.
We assume that the ratio of electron and ion currents is equal to $-T_e/T_i$.
To trigger magnetic reconnection, we locally pinch magnetic field lines with a perturbation $\mathbf{B}_1$ on the initial magnetic field (Eq.\ref{eq:B}):
\begin{equation}
	\textbf{B}_1 = B_{1x}(x,y) \textbf{u}_x + B_{1y}(x,y) \textbf{u}_y
    \label{eq:Bone}
\end{equation}    
\begin{equation}
  	 B_{1x}(x,y) = -2\delta b\frac{y-y_0}{\sigma}\exp{-\frac{(x-x_0)^2+(y-y_0)^2}{\sigma^2}}
   	\label{eq:Bonex}
\end{equation}    
\begin{equation}
     B_{1y}(x,y) =  2\delta b\frac{x-x_0}{\sigma}\exp{-\frac{(x-x_0)^2+(y-y_0)^2}{\sigma^2}}
	\label{eq:Boney}
\end{equation}
where $y_0 = y_{max}/2$, $x_0 = x_{max}/2$, $\delta b=0.12$ and $\sigma=1$.

We initialized our simulation with three ion species: the magnetosheath ions ($ish$), hot magnetospheric ions ($ih$) and plume ions ($ip$).
We only implement one population of electrons ($e$), neglecting the cold electrons which are not considered in our study.
We have $n = n_e = n_{ish}+n_{ih}+n_{ic}$.
The hot magnetospheric ions are reconnecting with magnetosheath ions before the impact of the plume.
They are tenuous compared to magnetosheath ions ($n_r=n_{ish}/n_{ih}=10$).
Their density is negligible compared to the plume's one, with a density ratio $n_{hop} = n_{ih}/n_{ip}=0.05$.
Their high temperature is essential for the pressure balance, as $T_{hop} = T_{ih}/T_{ip}=500$.
To calculate the density profile of each species, we make the assumption that each of these species has initially a constant temperature in the domain.
We determine the asymptotic densities thanks to the normalized pressure balance:
\begin{eqnarray}
	K & = & n_{ish}T_{ish} + n_{ih}T_{ih} +n_{ip}T_{ip} +n_{e}T_{e} + \frac{B^2}{2} \\
	   & = & \left( n_{ish}T_{ish} + n_{ih}T_{ih} +n_{ip}T_{ip} \right) (1 + \theta) + \frac{B^2}{2}
	\label{eq:pb}
\end{eqnarray}
where $K$ is a constant, fixed at $1/B_r^2 = 4$ in our case.

The ion temperatures being constant, we determine them by using asymptotic values of the density and applying \refe{eq:pb}.
Thus, we calculate from \refe{eq:pb}:
\begin{eqnarray}
	T_{ish} & = & \frac{K-1/2}{1+\theta} \\
	T_{ih}   & = & \frac{n_r}{1+\theta} \left( K - \frac{1}{2B_r^2}  \right)\\
	T_{ip}   & = & \frac{T_{ih}}{T_{hop}}
\end{eqnarray}
Note that to preserve both pressure balance and our previous assumptions of constant ions temperature, we can not keep $n_{ih} = 0.1$ in presence of the plume.
The asymptotic temperatures and densities for each population in the different area are summarized in \reff{tab:initasym}.

We fix the density profiles for the magnetosheath ions and the plume ions such as:
\begin{eqnarray}
	n_{ish}(x,y) & = & \frac{1}{2} \left[1+ \tanh \left(\frac{y-y_0}{L} \right) \right] \label{eq:nish} \\
	n_{ip}(x,y) & = & \left[1- \tanh \left(\frac{y-y_0+dy}{L} \right) \right] \label{eq:nip}
\end{eqnarray}
With these profiles, the temperatures and the \refe{eq:pb}, we determine the density for hot ions:
\begin{equation}
	n_{ih}(x,y) =  \frac{1}{T_{ih}} \left[ \frac{K-B(x,y)^2/2}{1+\theta} -n_{ish}(x,y)T_{ish} -n_{ip}(x,y)T_{ip} \right]
	\label{eq:ni}
\end{equation}

\bibliography{biblio}
\end{document}